\documentclass[aps,pra,amsmath,amssymb,reprint,groupedaddress]{revtex4-2}

\usepackage{epsfig,amsmath}
\usepackage{subfigure}
\usepackage{graphicx}
\usepackage{dcolumn}
\usepackage{stmaryrd}
\usepackage{mathrsfs}
\usepackage{pifont}
\usepackage{amsthm}
\usepackage{amssymb}
\usepackage{bm}
\usepackage{latexsym}
\usepackage[colorlinks=true,linkcolor=blue,citecolor=blue]{hyperref}
\usepackage{color}
\usepackage{epstopdf}
\usepackage{dsfont}
\usepackage{cases}
\usepackage{multirow}
\usepackage{extarrows}
\usepackage{booktabs}
\usepackage{mathtools}

\newcommand{\lam}{\lambda}

\newcommand{\de}{\delta}
\newcommand{\De}{\Delta}

\newcommand{\ti}{\tilde}
\newcommand{\la}{\langle}
\newcommand{\ra}{\rangle}
\newcommand{\ka}{\kappa}

\newcommand{\Ga}{\Gamma}
\newcommand{\om}{\omega}

\newcommand{\Si}{\Sigma}
\newcommand{\al}{\alpha}
\newcommand{\be}{\beta}

\newcommand{\vep}{\varepsilon}

\newcommand{\mP}{\mathcal{P}}

\newcommand{\hH}{\hat{H}}
\newcommand{\hQ}{\hat{Q}}
\newcommand{\hP}{\hat{P}}

\newcommand{\hI}{\hat{I}}
\newcommand{\hS}{\hat{S}}

\newcommand*{\dif}{\mathop{}\!\mathrm{d}}
\newcommand{\e}{\mathrm{e}}
\newcommand{\im}{\mathrm{i}}

\renewcommand{\Re}{\operatorname{Re}}
\renewcommand{\Im}{\operatorname{Im}}

\begin{document}

\title{Bound states and decay dynamics in $N$-level Friedrichs model with factorizable interactions}

\author{Jia-Ming Zhang}
\email{Contact author: zhangjiaming@sdust.edu.cn}

\author{Yu Xin}

\author{Bing Chen}

\affiliation{College of Electronics and Information Engineering, Shandong University of Science and Technology, Qingdao 266590, Shandong, China}

\date{\today}

\begin{abstract}
Considering an $N$-level system interacting factorizably with a continuous spectrum, we derive expressions for the bound states and the dynamical evolution within this single-excitation Friedrichs model by using the projection operator formalism. First, we establish explicit criteria to determine the number of bound states, whose existence suppresses the complete spontaneous decay of the system. Second, we derive the open system's decay dynamics, which is naturally described by an energy-independent non-Hermitian Hamiltonian in the Markovian limit. As an example, we apply our framework to a two-level atomic chain side-coupled to a photonic lattice, uncovering a rich variety of decay dynamics and realizing an anti-$\mathcal{PT}$-symmetric Hamiltonian in the system's evolution.
\end{abstract}

\maketitle

\section{Introduction}\label{intro}

Even a perfectly isolated quantum system inherently interacts with the electromagnetic vacuum~\cite{Cohen1998,Breuer2007}. The seminal work of Weisskopf and Wigner in the 1930s established that an isolated discrete state undergoes exponential decay into a continuum~\cite{Weisskopf1930}. Nevertheless, coherent superpositions of discrete states and the continuum can form dressed bound states outside the continuum (BOCs), whose number and energies are strongly influenced by the environmental properties~\cite{Miyamoto2005,Bulgakov2007,Cui2018}. Distinct from these conventional BOCs, the counterintuitive phenomenon of bound states in the continuum (BICs) demonstrates that robust localization can persist within a continuum, a feature that has been observed across diverse physical platforms~\cite{Hsu2016}.
Both BOCs and BICs, spanning the total Hilbert space, have enabled a variety of applications, such as protecting quantum entanglement~\cite{Bellomo2008,Lazarou2012,Facchi2016,Behzadi2018}, designing vortex lasers~\cite{Huang2020}, generating high harmonics~\cite{Koshelev2019,Koshelev2020}, enhancing sensing protocols~\cite{Liu2017}, and storing energy in quantum batteries~\cite{Song2024,Lu2025}.
Beyond bound-state physics, the dynamics of open quantum systems constitutes a distinct and active frontier of research. Recent decades have witnessed significant advances in understanding diverse dynamical phenomena, which often emerge from the intricate interplay between system-environment coupling and structured reservoirs. Prominent examples include fractional decay~\cite{Longhi2007}, dynamical phase transitions~\cite{Liang2025,Castillo2025},  non-Markovian decoherence~\cite{Sinayskiy2009,Gonzalez2017,Shen2019,Burgess2022}, and superradiant or subradiant emission of multiple emitters~\cite{Bin2022,Asselie2022,Cardenas2023,Masson2024,Han2024,Chu2025}. Understanding these rich dynamical features is crucial for controlling quantum coherence, designing novel photonic devices, and exploring fundamental limits in quantum thermodynamics.

Within the single-excitation subspace, the inevitable decay dynamics of an unstable quantum system is effectively captured by the original Friedrichs model~\cite{Friedrichs1948}. This minimal Hamiltonian framework describes a single discrete state coupled to a continuum, where the interaction transmutes the bound state into a resonance with a finite width.
Over the past few decades, it has been significantly extended to incorporate more complex scenarios~\cite{Antoniou2003,Ordonez2004,Courbage2007,Gadella2011,Lonigro2022,Xiao2017,Xiao2025}, and has found applications across a broad range of fields, encompassing quantum field theory~\cite{Araki1957}, equilibrium statistical mechanics~\cite{Bach2000}, quantum optics~\cite{Gadella2011}, and hadronic physics~\cite{Xiao2017,Xiao2025}.
For an $N$-level system coupled to a continuum with a lower energy bound, it is known that a BOC eigenenergy always exists below each discrete level outside the continuum, while an additional BOC eigenenergy may emerge from a discrete level inside the continuum~\cite{Miyamoto2005}. However, after eliminating the environmental degrees of freedom, the $N$-dimensional nonlinear equations governing these bound states are generally intractable except under weak coupling via perturbation theory or within a two-level model~\cite{Antoniou2003}.

In this work, we focus on a generalized Friedrichs model comprising $N$ discrete levels coupled to a continuum. Apart from the condition of a factorizable discrete-continuum interaction, we impose no restrictions on the structures of both the non-degenerate discrete system and the continuum. Under this factorizable assumption and within the projection operator formalism,  we derive explicit eigenvalue-dependent formal solutions for both bound and scattering states. We also propose simple and general criteria for determining the number of BOCs without numerically solving the implicit eigenvalue equation.
We further analyze the dynamical evolution of such an unstable system, with particular emphasis on its survival probability. It is shown that this dynamics is naturally described by an energy-independent non-Hermitian Hamiltonian in the Markovian limit. Finally, we illustrate the resulting decay phenomenology on a tight-binding model, which exhibits rich structural versatility~\cite{Dinc2019,Pivovarov2021,Longhi2025} and has been widely realized in experiments~\cite{Dreisow2008,Goban2015,Hood2016}. Within this setting, we demonstrate how an anti-$\mathcal{PT}$-symmetric Hamiltonian can be systematically constructed.

The paper is organized as follows. In Sec.~\ref{sec2}, we introduce the general model and briefly review the projection formalism along with the effective Hamiltonian approach. Section~\ref{BS} presents criteria for determining the number of bound states and provides expressions of these states. The scattering states and the survival probability dynamics are derived in Sec.~\ref{sec4}. In Sec.~\ref{sec:Mar}, we reveal how the open-system dynamics maps to a non-Hermitian description in the Markovian regime. Section~\ref{sec6} specializes the discussion to a two-level atomic chain side-coupled to a photonic lattice. Concluding remarks are given in Sec.~\ref{conc}.

\section{Basic model and effective Hamiltonian}\label{sec2}

The $N$-level Friedrichs model describes the interaction of an $N$-level discrete system with a continuum. Under the separable interaction condition, the total system Hermitian Hamiltonian is given by
\begin{equation}\label{H}
\begin{aligned}
\hH\!=\!{}&\hH_{\rm D}+\hH_{\rm C}+\hH_{\rm I}\\
=\!{}&\sum_{n=1}^N\vep_n |n\ra\la n|
+\int_{\om_{\rm low}}^{\om_{\rm up}} \om| \om\ra\la\om| \rho(\om) \dif\om\\
&\!+\!\sum_{n=1}^N \int_{\om_{\rm low}}^{\om_{\rm up}}\!\left[f_n g(\om)| n\ra\la\om| \!+\!f_n^* g^*(\om)| \om\ra\la n| \right]\!\sqrt{\rho(\om)}\dif\om,
\end{aligned}
\end{equation}
where $| n\ra$ and $| \om \ra$ are the bare energy eigenstates of the discrete system $\hH_{\rm D}$ with energies $\vep_n$ (such as atoms or quantum emitters), and of the continuum $\hH_{\rm C}$ with energies $\om$ (the environmental reservoir such as the electromagnetic field), respectively, in the absence of coupling between the two subsystems. These eigenstates satisfy $\la n| n'\ra=\de_{nn'}$, $\la \om| \om'\ra=\de(\om-\om')/\rho(\om)$, and $\la n| \om\ra=0$, where $\de_{nn'}$ is Kronecker's delta and $\de(\om-\om')$ is Dirac's delta function. $\rho(\om)$ is the density of the continuous states and the energy band allowed is a specific region $(\om_{\rm low},\om_{\rm up})$. It is assumed that $\vep_1<\vep_2<\cdots<\vep_N$ without any degeneracy, and the coupling strength characterizing the transition between $| n\ra$ and $| \om\ra$ can be factorized into two components $f_n$ and $g(\om)$. The asterisk denotes the complex conjugate.

The solution $| \Phi(E)\ra$ of the stationary Schr{\"o}dinger equation at energy $E$ satisfies
\begin{equation}\label{SSE}
(E-\hH)| \Phi(E)\ra=0.
\end{equation}
Since our attention is restricted to the single excitation in the discrete system, we introduce Feshbach's projection operators~\cite{Feshbach1962,Dittes2000,Rotter2009}
\begin{equation}
\hQ=\sum_{n=1}^N | n\ra\la n| \quad \text{and} \quad \hP=\int_{\om_{\rm low}}^{\om_{\rm up}} | \om\ra\la\om| \rho(\om)\dif\om
\end{equation}
to divide the solution $| \Phi(E)\ra$ into two components:  $\hQ| \Phi(E)\ra$ in the discrete system and $\hP| \Phi(E)\ra$ in the continuum. The operators follow $\hQ\hQ=\hQ$, $\hP\hP=\hP$, $\hQ\hP=\hP\hQ=0$ and $\hQ+\hP=\hat{I}$, where $\hat{I}$ is the identity matrix.
Operating $\hQ$ and $\hP$ separately on Eq.~(\ref{SSE}), we obtain two coupled equations,
\begin{subnumcases}{}
(E-\hQ\hH\hQ)\hQ| \Phi(E)\ra=(\hQ\hH\hP)\hP| \Phi(E)\ra,\label{SSE1}\\
(E-\hP\hH\hP)\hP| \Phi(E)\ra=(\hP\hH\hQ)\hQ| \Phi(E)\ra.\label{SSE2}
\end{subnumcases}
Considering the continuum eigenstates $| \om\ra$ satisfying $(\om-\hP\hH\hP)| \om\ra=0$,
we can solve Eq.~(\ref{SSE2}) as follows
\begin{equation}\label{PPhi}
{\hP| \Phi(E)\ra=}
\begin{dcases}
\tfrac{1}{E-\hP\hH\hP}(\hP\hH\hQ)\hQ| \Phi(E)\ra, &E\neq \om,\\
| \om\ra+\tfrac{1}{E-\hP\hH\hP}(\hP\hH\hQ)\hQ| \Phi(E)\ra, &E=\om.
\end{dcases}
\end{equation}
After substituting Eq.~(\ref{PPhi}) into Eq.~(\ref{SSE1}) to eliminate the continuum subspace, the state in the discrete system takes the form
\begin{subequations}
\begin{alignat}{2}
\left[E-\hH_{\rm eff}(E)\right]\hQ| \Phi(E)\ra =0,& \qquad &E\neq \om, &\label{EigenEq}\\
\left[E-\hH_{\rm eff}(E)\right]\hQ| \Phi(E)\ra =\hQ\hH\hP| \om\ra,& \qquad &E=\om,&\label{QPhi}
\end{alignat}
\end{subequations}

where the \emph{energy-dependent} effective Hamiltonian without any coupling or statistical approximation is
\begin{align}
\hH_{\rm eff}(E)={}&\hQ\hH\hQ-\hQ\hH\hP\frac{1}{E-\hP\hH\hP}\hP\hH\hQ \notag\\
={}&\sum_{n=1}^N\vep_n | n\ra\la n|  +\Si(E)\sum_{n,n'=1}^N f_n f_{n'}^*| n\ra\la n'|, \label{Heff}
\end{align}
with the self-energy
\begin{equation}
\Si(E)\equiv\int_{\om_{\rm low}}^{\om_{\rm up}} \frac{J(\om)}{E-\om}\dif\om \label{Si}
\end{equation}
and the spectral density
\begin{equation}
J(\om)\equiv| g(\om)| ^2\rho(\om).
\end{equation}
One can find that projecting the stationary Schr{\"o}dinger Eq.~(\ref{SSE}) onto the discrete subspace yields two distinct formalisms: the homogeneous Eq.~(\ref{EigenEq}) and the non-homogeneous Eq.~(\ref{QPhi}). The solutions
$| \Phi(E) \rangle$ of these equations correspond to the discrete bound states and the continuous scattering states of the total system, respectively. This distinction will be elaborated upon in the following discussion.

\section{Discrete bound states}\label{BS}
When the real energy $E$ is outside the continuum or satisfies $J(E)=0$ in the continuum, the energy-dependent operator $\hH_{\rm eff}(E)$ is a Hermitian Hamiltonian. The solution $E$ of the eigenvalue problem~(\ref{EigenEq}) is equivalent to the determinant of the matrix $E-\hH_{\rm eff}(E)$ being zero, which can be expressed as
\begin{equation}\label{SE}
\prod_{n=1}^N(E-\vep_n)-\Si(E)\sum_{n=1}^N | f_n|^2\prod_{\substack{n'=1\\n'\neq n}}^N(E-\vep_{n'})
=0,
\end{equation}
by invoking our factorizable-interaction hypothesis.
Hereafter, we substitute $| \Phi_m\ra$ for bound state $| \Phi(E_m)\ra$ with eigenvalue $E_m$. The total number $M$ of bound states both outside and inside the continuum will be discussed in the following.
\subsection{Bound states outside the continuum}\label{sec:BOC}

\begin{figure}[htbp]
\centering
\includegraphics[width=\linewidth]{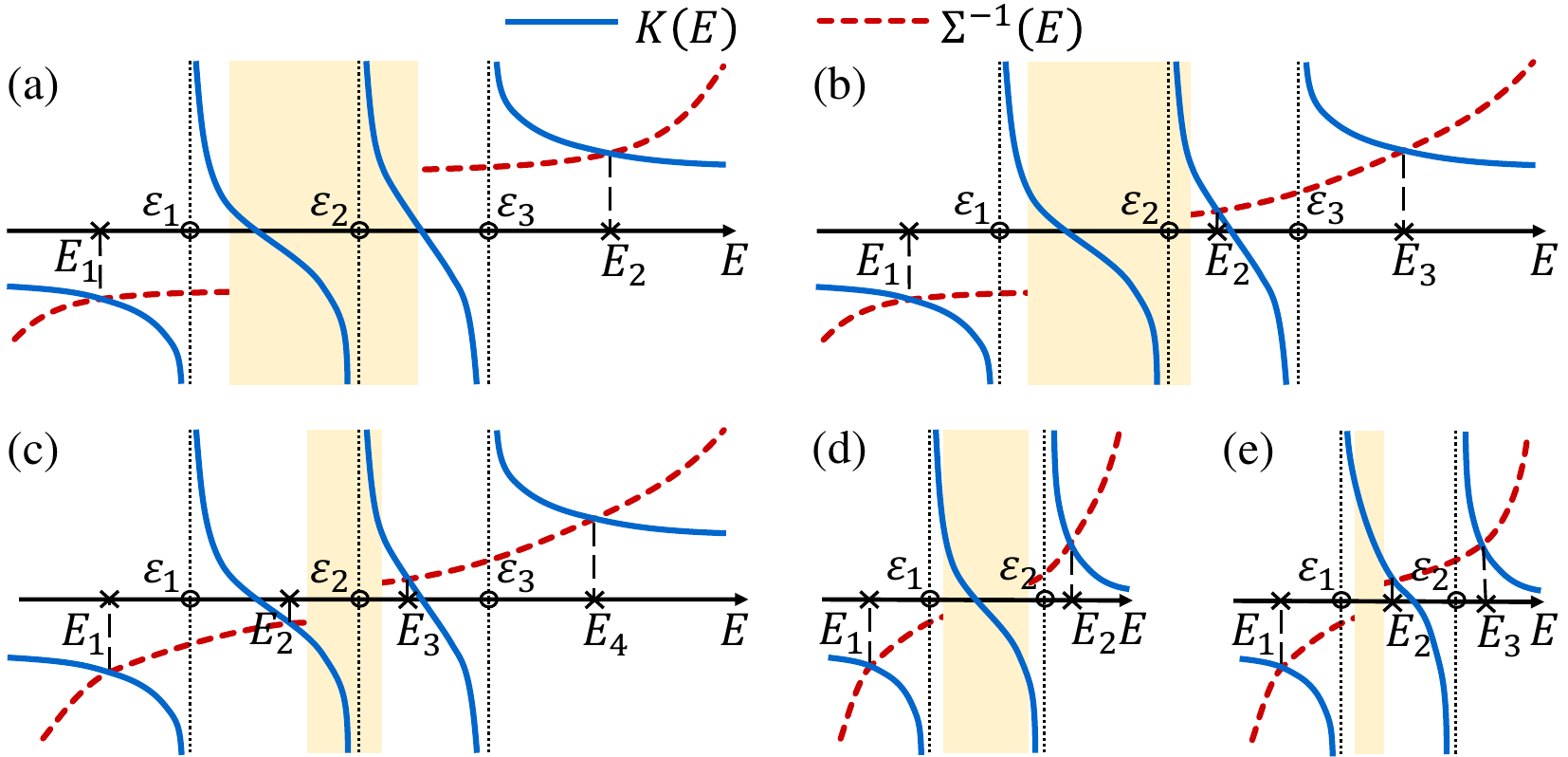}
\caption{Schematic graphical solution of Eq.~(\ref{sol}). The roots $E_m$ (black crosses) of Eq.~(\ref{sol}) are obtained from the intersection points between $K(E)$ (blue solid line) and $\Si^{-1}(E)$ (red dashed line). Black hollow circles and black dotted lines represent the energy level $\vep_n$ of the discrete system and the asymptotes of the function $K(E)$ at these positions, respectively. The light-yellow shaded region indicates the continuum  band. (a)--(c): The number of BOCs is $M_{\rm out}=N_{\rm out},N_{\rm out}+1, N_{\rm out}+2$, respectively, where $N_{\rm out}=2$ is the number of bare energies $\vep_n$ of the discrete system $\hH_{\rm D}$ lying outside the continuum. (d,e) The only two cases where all $N=2$ discrete levels are located outside the continuum, corresponding to $M_{\rm out}=N$ and $N+1$, respectively. In (e), only the situation with an additional bound state above the continuum is shown; the case with an additional bound state below the continuum is analogous.}
\label{schematic}
\end{figure}

The most common scenario is the presence of BOCs. With the assumption of no degeneracy among discrete levels $\{\vep_n\}_{n=1}^N$, Eq.~(\ref{SE}) can be rearranged. Dividing the equation by $\Si(E)\prod_{n=1}^N(E-\vep_n)$, we obtain a compact form
\begin{equation}\label{sol}
K(E)=\Si^{-1}(E),
\end{equation}
where we introduce the function
\begin{equation}
K(E)\equiv\sum_{n=1}^N\frac{ | f_n|^2}{E-\vep_n}.\label{K}
\end{equation}
Condition~(\ref{sol}) constitutes an implicit equation for $E$, which generally admits no algebraic solution. The graphical method of finding intersections between the curves $K(E)$ and $\Si^{-1}(E)$ offers a practical way to determine the existence of BOCs, as illustrated schematically in Fig.~\ref{schematic}.

The function $K(E)$ is strictly monotone decreasing, which follows directly from its derivative being negative, i.e.,
\begin{equation}\label{K'}
K'(E)=-\sum_{n=1}^N\frac{| f_n |^2}{(E-\vep_n)^2}<0.
\end{equation}
Combined with its asymptotic divergence at each discrete energy $\vep_n$, i.e., $\lim_{E\to\vep_n^\pm}K(E)=\pm\infty$, the function $K(E)$ must possess exactly $N-1$ real roots, with each root lying strictly between two consecutive discrete levels. Consequently, the plot of $K(E)$ consists of disjointed rather jagged branches. Traversing each interval $(\vep_n,\vep_{n+1})$ from left to right, $K(E)$ plummets from $+\infty$ at the left edge down to $-\infty$ at the right edge. As $E\to\pm\infty$, $K(E)$ approaches $0$ from above and below, respectively.
However, the properties of $\Si^{-1}(E)$ are comparatively straightforward. Given that
\begin{equation}
\Si'(E)=-\int_{\om_{\rm low}}^{\om_{\rm up}} \frac{J(\om)}{(E-\om)^2}\dif\om<0,
\end{equation}
its derivative satisfies
\begin{equation}
\frac{\dif \Si^{-1}(E)}{\dif E}=-\frac{\Si'(E)}{\Si^2(E)}>0.
\end{equation}
Outside the band, $\Si^{-1}(E)$ increases monotonically from $\lim_{E\to -\infty}\Si^{-1}(E)=-\infty$ to $\Si^{-1}(\om_{\rm low})\le 0$ on the low-energy side, and from $\Si^{-1}(\om_{\rm up})\ge 0$ to $\lim_{E\to \infty}\Si^{-1}(E)=+\infty$ on the high-energy side.

\begin{table}[htbp]
\caption{Number of BOCs.}
\label{table}
\begin{ruledtabular}
\renewcommand{\arraystretch}{1.5}
\begin{tabular}{ccc}
Energy region & Criterion & Number of BOCs \\
\hline
Below the band
& $K(\omega_{\mathrm{low}}) \le \Sigma^{-1}(\omega_{\mathrm{low}})$ & $N_{\mathrm{low}} + 1$ \\
\cmidrule{2-3}
$E \le \omega_{\mathrm{low}}$ & $K(\omega_{\mathrm{low}}) > \Sigma^{-1}(\omega_{\mathrm{low}})$ & $N_{\mathrm{low}}$ \\[4pt]
\hline
Above the band
& $K(\omega_{\mathrm{up}}) \ge \Sigma^{-1}(\omega_{\mathrm{up}})$ & $N_{\mathrm{up}} + 1$ \\
\cmidrule{2-3}
$E \ge \omega_{\mathrm{up}}$ & $K(\omega_{\mathrm{up}}) < \Sigma^{-1}(\omega_{\mathrm{up}})$ & $N_{\mathrm{up}}$ \\
\end{tabular}
\end{ruledtabular}
\end{table}

Suppose that there are $N_{\rm out}=N_{\rm low}+N_{\rm up}$ bare energies of the discrete system Hamiltonian $\hH_{\rm D}$ outside the band, where $N_{\rm low}$ and $N_{\rm up}$ are the numbers of discrete levels located in the domain $E\le\om_{\rm low}$ and $E\ge\om_{\rm up}$, respectively. If the spectral density $J(\om)$ vanishes as a power law near the band edges, i.e., $J(\om)\sim(\om-\om_{\rm low})^{s_{\rm low}}$ as $\om\to\om_{\rm low}$ and $J(\om)\sim(\om_{\rm up}-\om)^{s_{\rm up}}$ as $\om\to\om_{\rm up}$ with exponents $s_{\rm low},s_{\rm up}>0$, the self-energy $\Si(\om)$ defined in Eq.~(\ref{Si}) is convergent at the edges. Otherwise, $\Si(\om_{\rm low})$ and $\Si(\om_{\rm up})$ diverge, with $\Si^{-1}(\om_{\rm low})\to 0^-$ and $\Si^{-1}(\om_{\rm up})\to 0^+$.
For energies below the continuum ($E \le \omega_{\rm low}$), if $K(\omega_{\rm low}) > \Sigma^{-1}(\omega_{\rm low})$, the curves $K(E)$ and $\Sigma^{-1}(E)$ intersect at $N_{\rm low}$ points, yielding $N_{\rm low}$ solutions to Eq.~(\ref{sol}), each below a discrete level [see Fig.~\ref{schematic}(a) and \ref{schematic}(b)]. Conversely, if $K(\omega_{\rm low}) \le \Sigma^{-1}(\omega_{\rm low})$, one additional solution emerges above the highest discrete level that lies below the continuum [see Fig.~\ref{schematic}(c)].
Analogously, for energies above the continuum ($E \ge \omega_{\rm up}$), if $K(\omega_{\rm up}) < \Sigma^{-1}(\omega_{\rm up})$, the curves intersect at $N_{\rm up}$ points, yielding $N_{\rm up}$ solutions to Eq.~(\ref{sol}), each above a discrete level [see Fig.~\ref{schematic}(a)]. Conversely, if $K(\omega_{\rm up}) \ge \Sigma^{-1}(\omega_{\rm up})$, one additional solution emerges below the lowest discrete level that lies above the continuum [see Fig.~\ref{schematic}(b) and \ref{schematic}(c)].
For clarity, these graphical criteria and the corresponding number of BOCs are summarized in Table~\ref{table}.

Notably, if all $N$ discrete levels are located outside the continuum, the total number of BOCs is either $N$ or $N+1$, shown in Fig.~\ref{schematic}~(d) and (e), and $N+2$ is excluded. Indeed, the simultaneous satisfaction of $K(\omega_{\mathrm{low}}) \le \Sigma^{-1}(\omega_{\mathrm{low}})$ and $K(\omega_{\mathrm{up}}) \ge \Sigma^{-1}(\omega_{\mathrm{up}})$, which would be required for $N+2$, is impossible here.
In addition, the upper edge of the continuum may extend to infinity in many cases~\cite{Weiss2012}. Given this situation, no bare energies of $\hH_{\rm D}$ lie above the continuum, and BOCs can only appear below the continuum.

The eigene Eq.~(\ref{EigenEq}) implies that the projection of the BOC onto the discrete subspace, $\hQ| \Phi_m\rangle$, can be expanded in the eigenstates $\{| n \ra\}_{n=1}^N$ of the discrete system Hamiltonian as
\begin{equation}\label{Qm1}
\hQ| \Phi_m\ra=B_m\sum_{n=1}^N \frac{f_n}{E_m-\vep_n} | n\ra.
\end{equation}
We can also derive the projection of the BOC onto the continuum from Eq.~(\ref{PPhi}),
\begin{equation}\label{Pm1}
\hP| \Phi_m\ra=B_m\sum_{n=1}^N \frac{| f_n |^2}{E_m-\vep_n}\int_{\om_{\rm low}}^{\om_{\rm up}}\frac{g^*(\om)\sqrt{\rho(\om)}}{E_m-\om}| \om\ra\dif\om.
\end{equation}
The normalization constant $B_m$ used above are defined as
\begin{equation}\label{bm1}
| B_m| ^2
=-\frac{\Si(E_m)}{K'(E_m)\Si(E_m)+K(E_m)\Si'(E_m)},
\end{equation}
which is obtained from the normalization condition $\la\Phi_m| \hQ+\hP| \Phi_m\ra=1$.

\subsection{Bound states inside the continuum}\label{sec:BIC}
The origin of BICs lies in destructive interference in some parts of the continuum,  either among waves emitted from different discrete states or between an emitted wave and its reflection from the continuum. As reported in Refs.~\cite{Miyamoto2005} and~\cite{Longhi2007}, the existence of a BIC at frequency $E_m$ mathematically requires that both $J(E_m) = 0$ and Eq.~(\ref{SE}) hold simultaneously.
The former condition indicates that a point-like gap exists in the density of states $\rho(E_m)$ inside the band~\cite{Lambropoulos2000}, or one or more discrete states do not interact with the continuous state of energy $E_m$, e.g., $g(E_m)=0$. The latter condition constrains the corresponding energies to specific eigenvalues.

The expressions of the BICs are identical in form to Eqs.~(\ref{Qm1}) and (\ref{Pm1}). A special case arises when $\Si(E_m)=0$,  where state $| m\ra$ is also the eigenstate of $\hH_{\rm eff}(E_m)$, i.e.,
\begin{equation}\label{Qm2}
\hQ| \Phi_m\ra=B_m| m\ra.
\end{equation}
Analogous to previous analysis, the normalization constant satisfies
\begin{equation}\label{bm2}
| B_m| ^2=\frac{1}{1-| f_m |^2\Si'(E_m)},
\end{equation}
by normalizing the total wavefunction with
\begin{equation}
\hP| \Phi_m\ra= B_m f_m^*\int_{\om_{\rm low}}^{\om_{\rm up}}\frac{g^*(\om)\sqrt{\rho(\om)}}{E_m-\om}| \om\ra\dif\om.
\end{equation}

\section{Decay dynamics}\label{sec4}

\subsection{Continuous scattering states}
When eigenenergy $E$ coincides with the continuous spectrum, the effective Hamiltonian~(\ref{Heff}) is not well defined since the integrand in $\Si(E)$ has a pole on the real axis. It has to be considered as a limiting value from the upper or lower half of the complex energy plane, $E^\pm=E\pm\im\eta$, if the outgoing or incoming wave boundary condition is adopted. For $E$ belonging to the continuum spectrum, we define the non-Hermitian Hamiltonian~\cite{Rotter2009,Moiseyev2009,Moiseyev2011}
\begin{align}
\hH_{\rm eff}^\pm(E)\equiv&\lim_{\eta\to 0}\hH_{\rm eff}(E\pm\im\eta) \notag\\
=&\sum\limits_{n=1}^N\vep_n | n\ra\la n|  +\Si^\pm(E)\sum\limits_{n,n'=1}^N f_n f_{n'}^*| n\ra\la n'|,\label{Heff+-}
\end{align}
with the self-energy rewritten as
\begin{equation}\label{Si+}
\Si^\pm(E)\equiv\lim_{\eta\to 0^+}\Si(E\pm\im\eta)
=\De(E)\mp\im\Ga(E).
\end{equation}
Here the real functions $\De(E)$ and $\Ga(E)$ are obtained by applying the well-known Sokhotski-Plemelj formula, $\lim_{\eta\to 0^+}1/(x\pm\im\eta)=\mP\left(1/x\right)\mp\im\pi\de(x)$, which yields the energy shift
\begin{equation}\label{De}
\De(E)=\mP\int_{\om_{\rm low}}^{\om_{\rm up}} \frac{J(\om)}{E-\om}\dif\om,
\end{equation}
with $\mP$ denoting the Cauchy principal value, and the resonance width
\begin{equation}\label{Ga}
\Ga(E)=\pi J(E).
\end{equation}
With definition~(\ref{Heff+-}), Eq.~(\ref{QPhi}) reduces precisely to the Lippmann-Schwinger equation projected onto the discrete subspace.

To distinguish it from the bound states discussed in Sec.~\ref{BS}, we denote the outgoing scattering state by $| \Phi^+(E)\ra$, normalized as $\la\Phi^+(E')| \Phi^+(E)\ra=\de(E-E')$. From Eq.~(\ref{QPhi}), the projection of the scattering state on to the discrete system modified by its interaction with the continuum can be expanded in the basis $\{| n \ra\}_{n=1}^N$,
\begin{equation}\label{QPhi+}
\begin{aligned}
\hQ| \Phi^+(E)\ra\!={}&\!\frac{g(E)\sqrt{\rho(E)}}{E-\hH_{\rm eff}^+(E)}\sum_{n=1}^N f_{n}| n\ra\\
={}&\!\frac{g(E)\sqrt{\rho(E)}}{1\!-\!\De(E)K(E)\!+\!\im\Ga(E)K(E)}\!\sum_{n=1}^N \frac{f_n}{E-\vep_n}| n\ra.
\end{aligned}
\end{equation}

\subsection{Survival probability}

Once the bound states and scattering states are obtained, the time evolution of the projected state $\hQ| \phi(t)\ra$ can be decomposed into two parts,
\begin{align}
\hQ| \phi(t)\ra=&\sum_{m=1}^M \la\Phi_m| \phi(0)\ra\e^{-\im E_m t}\hQ| \Phi_m\ra \notag\\
&+\!\int_{\om_{\rm low}}^{\om_{\rm up}}\dif E\ \la\Phi^+(E)| \phi(0)\ra\e^{-\im E t}\hQ| \Phi^+(E)\ra.\label{qt}
\end{align}
We consider the decay dynamics of the most general initial state describing a single excitation prepared in the discrete system. This state allows for an arbitrary population distribution among the discrete states and is written as a superposition of the discrete energy eigenstates,
\begin{equation}
| \phi(0)\ra=\sum_{n=1}^N c_n | n\ra,
\end{equation}
with arbitrary complex coefficients $c_n$ satisfying $\sum_{n=1}^N | c_n|^2=1$. The decay function $p(t)$ is defined as the survival probability of finding the excitation in the discrete system at time $t$, namely,
\begin{align}
p(t)=&\sum_{n=1}^N \left| \la n| \hQ| \phi(t)\ra\right| ^2 \notag\\
=&\sum_{n=1}^N \left| \sum_{m=1}^M R_{nm}\e^{-\im E_m t}+\int_{\om_{\rm low}}^{\om_{\rm up}}\dif E\ S_{nm}\e^{-\im E t}\right| ^2 \label{pt}
\end{align}
with $p(0)=1$ and $p(t)\leq 1$ for any $t\geq 0$. Based on Eqs.~(\ref{Qm1}), (\ref{bm1}) to (\ref{bm2}), and (\ref{QPhi+}), we have
\begin{equation}
R_{nm}=\begin{dcases}
-\tfrac{1}{K'(E_m)+K^2(E_m)\Si'(E_m)}\frac{f_n I(E_m)}{E_m-\vep_n},&E_m\neq  \vep_n,\\
\tfrac{c_n\de_{mn}}{1-| f_n|^2\Si'(E_m)}, &E_m=\vep_n,
\end{dcases}
\end{equation}
and
\begin{equation}
S_{nm}=\frac{\Ga(E)}{\pi\{[1-\De(E)K(E)]^2+[\Ga(E)K(E)]^2\}}\frac{f_n I(E) }{E-\vep_n},
\end{equation}
with the function
\begin{equation}\label{I}
I(E)=\sum_{n=1}^N \frac{f_n^* c_n}{E-\vep_n}.
\end{equation}
It is worth noting that the integral in Eq.~(\ref{pt}) vanishes as $t$ goes to infinity, as a consequence of the Riemann-Lebesgue lemma~\cite{Carl1999}. In the long time limit, the survival probability reduces to the contribution from the bound states,
\begin{align}
P(t)\equiv&\lim_{t\to\infty}p(t)
=\sum_{n=1}^N \left|\sum_{m=1}^M R_{nm}\e^{-\im E_m t}\right| ^2 \notag\\
=&\sum_{m=1}^M \sum_{n=1}^N | R_{nm} |^2 + \sum_{m>m',m'=1}^M O_{mm'}(t),
\end{align}
where the first term is time-independent and gives the long-time average, while the second contains all oscillatory contributions
\begin{equation}
\begin{aligned}
O_{mm'}(t)={}&2\left|\sum_{n=1}^N R_{nm}R_{nm'}^*\right|\\
&\!\times\!\cos\left[(E_m\!-\!E_{m'})t
\!-\!\arg\left(\sum_{n=1}^N R_{nm}R_{nm'}^*\right)\right],
\end{aligned}
\end{equation}
with frequencies $E_m-E_{m'}$.
The oscillatory terms vanish if the cross-coefficients satisfy $\sum_{n=1}^N R_{nm}R_{nm'}^*=0$. This condition is realised not only for a single bound state ($M=1$), but also for multiple bound states, provided that all of them are BICs with zero self-energy. In the multiple BICs case, each BIC corresponds to a distinct initial eigenstate of the discrete system, as shown in Eq.~(\ref{Qm2}), such that the coefficients $R_{nm}$ are diagonal. Consequently, different BICs do not overlap and no quantum interference occurs in the discrete subspace. The asymptotic survival probability therefore becomes a constant when either only one bound state exists, or when the entire set of bound states consists solely of BICs with zero self-energy; it decays to zero if no bound states exist, and oscillates otherwise.

\section{Markovian regime}\label{sec:Mar}
Under the Weisskopf-Wigner or Markovian approximation, the continuum is nonstructured, which leads to an infinite bandwidth with a constant spectral density $J(\om) = J$.
In this limit, there are no band edges and $J(E)$ is nowhere zero, so neither BOCs nor BICs can exist. The absence of these bound states leads to completely dissipative dynamics.
In addition, the energy shift $\De(E)$ in Eq.~(\ref{De}) vanishes, and the effective Hamiltonian~(\ref{Heff+-}) becomes energy-independent,
\begin{equation}\label{HeffMar}
\hH_{\rm eff}^\pm (E)=\hH_{\rm eff}^\pm=\sum_{n=1}^N\vep_n | n\ra\la n|  +\Si^\pm\sum_{n,n'=1}^N f_n f_{n'}^*| n\ra\la n'|,
\end{equation}
with $\Si^\pm=\mp\im \Ga$, and $\Ga=\pi J$.
According to Eq.~(\ref{QPhi+}), Eq.~(\ref{qt}) collapses into a more concise form
\begin{equation}\label{Qt}
\hQ|\phi(t)\ra
=\e^{-\im \hH_{\rm eff}^+ t} |\phi(0)\ra.
\end{equation}
Consequently, the evolution is characterized by the exponential operator of the non-Hermitian effective Hamiltonian $\hH_{\rm eff}^+$.

Following standard techniques for non-Hermitian operators, let us indicate in the discrete subspace the eigenstates of $\hH_{\rm eff}^\pm$ by $\{| \Psi_i^\pm\ra\}_{i=1}^N$, which form a bi-orthogonal basis of the Hilbert space and its dual space. These states are resonance states of the open system, characterized by their finite lifetimes. When such eigenstates are non-degenerate, we adopt the orthonormality conditions or c-product~\cite{Moiseyev2011,Brody2014}
\begin{equation}
\la\Psi_{i'}^-| \Psi_i^+\ra=\de_{ii'},
\end{equation}
satisfying the closure relation
\begin{equation}
\sum_{i=1}^N| \Psi_i^+\ra\la\Psi_i^-| =\hat{I}.
\end{equation}
Thus, the projected state $\hQ| \phi(t)\ra$ in Eq.~(\ref{Qt}) can be expressed as
\begin{equation}
\hQ| \phi(t)\ra=\sum_{i=1}^N \la\Psi_i^-| \phi(0)\ra\e^{-\im z_i t}| \Psi_i^+\ra.
\end{equation}

Using the distinct eigenvalue $z_i$ of $\hH_{\rm eff}^+$ that possesses negative imaginary part, the corresponding right eigenstates $| \Psi_i^\pm\ra$ can be expanded on the orthonormal and complete basis $\{| n\ra\}_{n=1}^N$ as
\begin{equation}
| \Psi_i^+\ra=V_i\sum_{n=1}^N \frac{f_n}{z_i-\vep_n}| n\ra, \quad
| \Psi_i^-\ra=W_i\sum_{n=1}^N \frac{f_n}{z_i^*-\vep_n}| n\ra,
\end{equation}
with the normalization functions satisfying
\begin{equation}
V_i W_i^*=-\frac{1}{K'(z_i)}.
\end{equation}
After ordering the eigenvalues such that $\Im z_N <\cdots<\Im z_2<\Im z_1<0$, we rewrite the decay function~(\ref{pt}) as
\begin{align}
p(t)={}&D_1\e^{2\Im (z_1) t}\Bigg[1+\sum_{i=2}^N\frac{D_i}{D_1}\e^{2(\Im z_i-\Im z_1)t}\notag\\
&+2\sum_{i>i'}^N \frac{U_{ii'}(t)}{D_1}\e^{(\Im z_i+\Im z_{i'}-2\Im z_1)t}\Bigg],\label{decay}
\end{align}
with the time-independent function
\begin{equation}
D_i=\sum_{n=1}^N \left|\frac{f_n I(z_i)}{(z_i-\vep_n)K'(z_i)}\right|^2
\end{equation}
and the time-dependent function
\begin{widetext}
\begin{equation}
U_{ii'}(t)=\left|\sum_{n=1}^N \frac{| f_n |^2 I(z_i)I(z_{i'}^*)}{K'(z_i)K'(z_{i'}^*)(z_i-\vep_n)(z_{i'}^*-\vep_n)} \right|
\cos\left[(\Re z_i-\Re z_{i'})t-\arg\left(\sum_{n=1}^N \frac{| f_n |^2 I(z_i)I(z_{i'}^*)}{K'(z_i)K'(z_{i'}^*)(z_i-\vep_n)(z_{i'}^*-\vep_n)} \right)\right].
\end{equation}
\end{widetext}
As Eq.~(\ref{decay}) shows, the survival probability exhibits a non-exponential decay due to multi-resonance interference. The real and imaginary parts of the resonance eigenvalues determinate the oscillation frequencies and the decay rates, respectively. At long times, the excitation decays exponentially with a rate $2| \Im z_1|$ .

If $\hH_{\rm eff}^+$ in Eq.~(\ref{HeffMar}) is a defective matrix, at least two complex eigenvalues will cross at an exceptional point (EP). For the sake of simplicity, let us assume that all $N$ eigenstates of $\hH_{\rm eff}^+$ coalesce. We indicate the eigenstate of $\hH_{\rm eff}^+$ by $| \Psi_{{\rm d},1}^+\ra$ with eigenvalue $z_{\rm d}$, \begin{equation}
(\hH_{\rm eff}^+ -z_{\rm d})| \Psi_{{\rm d},1}^+\ra=0,
\end{equation}
and by $\{| \Psi_{{\rm d},j}^+\ra\}_{j=2}^N$ the chain of associated eigenvectors,
\begin{equation}
(\hH_{\rm eff}^+ -z_{\rm d})| \Psi_{{\rm d},j}^+\ra=| \Psi_{{\rm d},j-1}^+\ra.
\end{equation}
$\{| \Psi_{{\rm d},j}^+\ra\}_{j=1}^N$ are linearly independent, and $| \Psi_{{\rm d},1}^+\ra$ is self-orthogonality~\cite{Moiseyev2011}. The closure relations for the $N$-dimensional space are given by
\begin{equation}
\hI=\sum_{j,j'=1}^N (\hS^{-1})_{j,j'}| \Psi_{{\rm d},j}^+\ra \la\Psi_{{\rm d},j'}^-|,
\end{equation}
with $\hS_{j,j'}=\la\Psi_{{\rm d},j}^-|\Psi_{{\rm d},j'}^+\ra$. Therefore, we arrive at
\begin{equation}
\begin{aligned}
\hQ| \phi(t)\ra={}&
\e^{-\im z_{\rm d} t}\sum_{j,j'=1}^N\Bigg\{\la\Psi_{{\rm d},j'}^-| \phi(0)\ra\\
&\times\left[\sum_{k=0}^{N-j}
\frac{\left(-\im t\right)^k}{k!}(\hS^{-1})_{k+j,j'}\right]| \Psi_{{\rm d},j}^+\ra\Bigg\},
\end{aligned}
\end{equation}
and the decay function yields
\begin{equation}\label{ddcay}
\begin{aligned}
p(t)\!=\!{}&\e^{2\Im (z_{\rm d}) t}\sum_{n=1}^N \Bigg| \la n | \phi(0) \ra\\
&\!+\!\sum_{j,j'=1}^N\!\la n| \Psi_{{\rm d},j}^+\ra\la\Psi_{{\rm d},j'}^-|\phi(0)\ra\!\left[\sum_{k=1}^{N\!-\!j}
\frac{\left(\!-\!\im t\right)^k}{k!}(\hS^{\!-\!1})_{k\!+\!j,j'}\right]\!\Bigg| ^2\!.
\end{aligned}
\end{equation}
Clearly, in the asymptotic limit $t\to\infty$, $p(t)$ shows a power-law exponential decay $\sim t^{2(N-1)}\e^{2\Im (z_{\rm d}) t}$, which was discussed in detail in Ref.~\cite{Longhi2018}.

\section{Explicit example: a two-level atomic chain side-coupled to a photonic lattice}
\label{sec6}

\begin{figure}[htbp]
\centering
\subfigure{
    \includegraphics[width=0.45\linewidth]{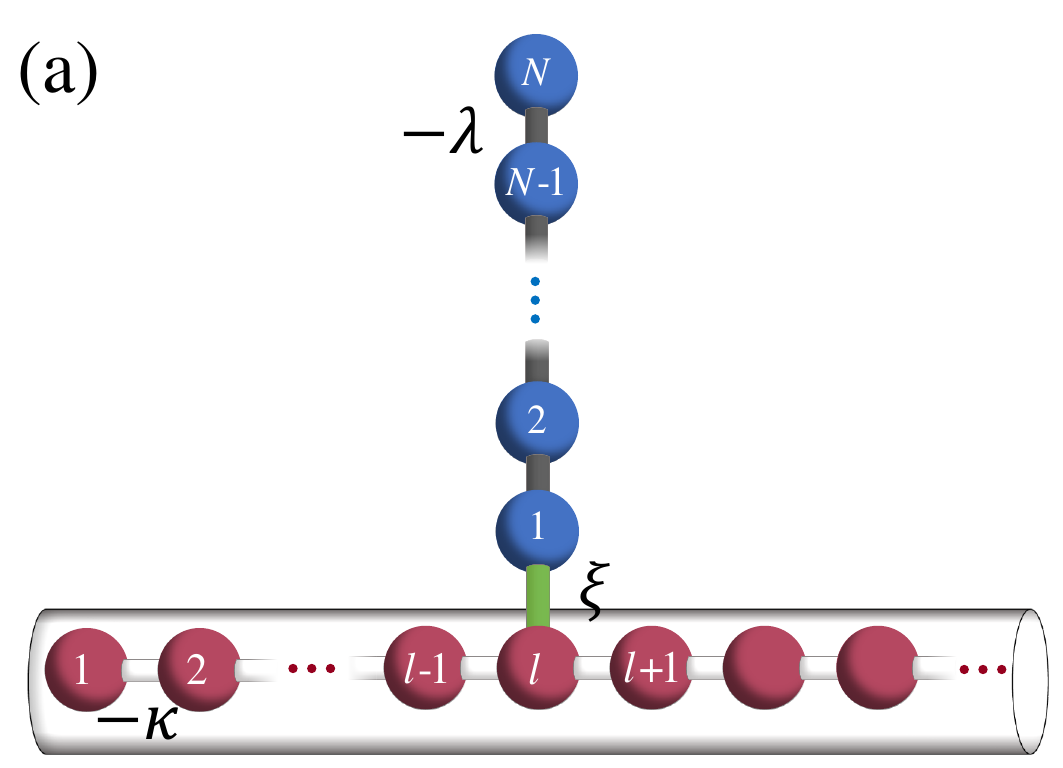}
    \label{TBM}
}
\subfigure{
    \includegraphics[width=0.45\linewidth]{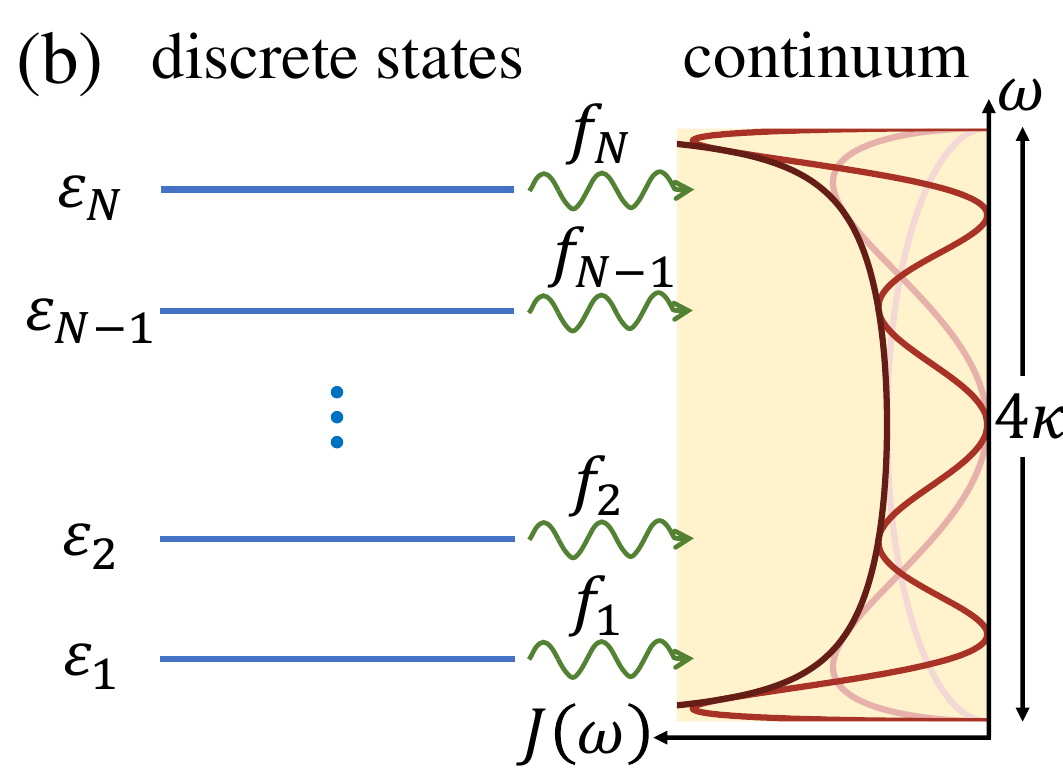}
    \label{Bloch}
}
\caption{Schematic of a two-level atomic chain coupled to a photonic lattice. (a) Tight-binding model in the Wannier basis. The system consists an $N$-site chain of two-level atoms, which is side-coupled to a one-dimensional semi-infinite photonic lattice at site $l$. (b) Corresponding energy spectrum. The spectral densities $J(\om)$ are shown by the red curves, with colors ranging from light to dark corresponding to $l=1$, $l=2$, $l=4$, and the asymptotic limit $l\to\infty$.}
\end{figure}

Let us consider an idealized system consisting of a chain of two-level artificial atoms (e.g., quantum dots or superconducting qubits) side-coupled to a semi-infinite tight-binding photonic lattice, as schematically depicted in Fig.~\ref{TBM}. This system, which has been used to model multilevel decay~\cite{Longhi2018}, can be realized with atom-like emitters coupled to a photonic crystal waveguide~\cite{Yariv1999}, a quantum wire~\cite{Orellana2006,Perczel2020}, an array of waveguides~\cite{Dreisow2008}, or an array of Josephson-junction resonators~\cite{Scigliuzzo2022}. The dynamics of one or two atoms coupled to a tight-binding photonic lattice has been extensively studied in the literature~\cite{Longhi2006,Dente2008,Calajo2019,Garmon2019}. Meanwhile, the bound states in such a system, known as the atom-photon bound states, have also attracted considerable attention recently~\cite{Castillo2025,Longhi2025,Scigliuzzo2022}. In practice, dissipation and various imperfections, such as radiative leakage into free-space modes, imperfect confinement of waveguide modes, and fabrication-induced disorder, are unavoidable. These effects can substantially reduce the propagation length of photons and alter the bound-state dynamics.
However, taking the platform of artificial atoms embedded in an array of compact superconducting resonators as an example, the free-space emission rate ($\sim$50\,kHz) and the resonator leakage rate ($\sim$300\,kHz) are much smaller than both the atom-array coupling strength ($\sim$300\,MHz) and the nearest-neighbor coupling between resonators ($\sim$250\,MHz)~\cite{Scigliuzzo2022}. Fabrication imperfections affect individual modes, but the collective waveguide behavior remains close to ideal due to the strong resonator-resonator coupling.
To present the main physics of the system, we neglect these additional dissipation and nonuniform couplings.

The total Hamiltonian of the ideal system is
\begin{align}
\hH={}&\hH_{\rm D}+\hH_{\rm C}+\hH_{\rm I} \notag\\
={}&-\lam\sum_{\mu=1}^{N-1}\left(| \mu\ra_{\rm a}{}_{\rm a}\la \mu+1| +| \mu+1\ra_{\rm a}{}_{\rm a}\la \mu| \right) \notag\\
{}&-\ka\sum_{\nu=1}^\infty \left(| \nu\ra_{\rm p}{}_{\rm p}\la \nu+1| +| \nu+1\ra_{\rm p}{}_{\rm p}\la \nu| \right) \notag\\
&+\xi\left(| 1\ra_{\rm a}{}_{\rm p}\la l | +| l\ra_{\rm p}{}_{\rm a}\la 1| \right).\label{TBMH}
\end{align}
Here the chain of two-level artificial atoms is described by $N$ Wannier sites with hopping amplitude $\lam$. The semi-infinite photonic lattice is described by a one-dimensional tight-binding Hamiltonian with hopping amplitude $\ka$ in the Wannier representation. $| \mu\ra_{\rm a}$ and $| \nu\ra_{\rm p}$ are states representing the single-excitation at the $\mu$th atom and $\nu$th photonic lattice site, respectively. It is assumed that one end of the atomic chain labeled $1$ couples to the $l$th site of the photonic lattice with strength $\xi$. The excitation is initially prepared at the open end of the chain, e.g.,
\begin{equation}\label{ini}
|\phi(0)\ra=| N\ra_{\rm a}.
\end{equation}

To map the tight-binding model onto the general Friedrichs model, we introduce the Bloch states $|n\ra$ and $|k\ra$ of the atomic chain and the photonic lattice, which are decoupled from each other. Their relations to the corresponding Wannier sites $|\mu\ra_{\rm a}$ and $|\nu\ra_{\rm p}$ are given by~\cite{Longhi2007,Longhi2009,Longhi2018}
\begin{align}
| n\ra={}&\sqrt{\frac{2}{N+1}}\sum_{\mu=1}^N\sin\frac{\pi n\mu}{N+1}| \mu\ra_{\rm a}, && n = 1, 2, \ldots, N, \notag\\
| k\ra={}&\sqrt{\frac{2}{\pi}}\sum_{\nu=1}^\infty \sin (k\nu)| \nu\ra_{\rm p}, && k \in (0, \pi).
\end{align}
Here $|n\ra$ are bare energy eigenstates of the uncoupled atomic chain, with the non-degenerate energy levels
\begin{equation}
\vep_n=-2\lam\cos\frac{\pi n}{N+1}.
\end{equation}
After introducing the density of continuous states given by
\begin{equation}
\rho(\om)=\frac{\partial k}{\partial\om}=\frac{\Theta(4\ka^2-\om^2)}{\sqrt{4\ka^2-\om^2}},
\end{equation}
the momentum eigenstates $|k\ra$  of the photonic lattice are converted into energy eigenstates $|\om\ra$ via the dispersion relation
\begin{equation}\label{dispersion}
\om(k)=-2\ka\cos k.
\end{equation}
We then immediately obtain an $N$-level Friedrichs model described by Eq.~(\ref{H}), with the separable coupling functions given by
\begin{align}
f_n={}&\xi\sqrt{\frac{2}{N+1}}\sin\frac{\pi n}{N+1}, \notag\\
g(\om)={}&\sqrt{\frac{2}{\pi}}\sin\left[l\arccos\left(-\frac{\om}{2\ka}\right)\right],
\end{align}
and the initial state defined by Eq.~(\ref{ini}) takes the form
\begin{equation}
| \phi(0)\ra=\sqrt{\frac{2}{N+1}}\sum_{n=1}^N\sin\frac{\pi n N}{N+1}| n\ra
\end{equation}
in the energy eigenbasis.
For $l\ge 2$, the spectral density
\begin{equation}\label{Jw}
J(\om)=\frac{2\Theta(4\ka^2-\om^2)}{\pi\sqrt{4\ka^2-\om^2}} \sin^2\left(l\arccos\frac{\om}{2\ka}\right)
\end{equation}
oscillates and has $l-1$ zeros at $\om=-2\ka\cos(\pi q/l)$ for $q=1,2,\ldots,l-1$, while vanishing at the band edges, as shown in Fig.~\ref{Bloch}.

Utilizing the residue theorem, the self-energy in Eq.~(\ref{Si}) can be evaluated exactly, yielding the expression
\begin{equation}
\Si(z)=-\im\dfrac{1-\exp\left[\im 2 l \arccos\left(-\frac{z}{2\ka}\right)\right]}{2\ka\sin\left[ \arccos\left(-\frac{z}{2\ka}\right)\right]},
\end{equation}
by analytically continuing the energy $E$ to the complex variable $z$~\cite{Abramowitz1965}. For a real energy $E$, $\Si(E)$ takes the piecewise form
\begin{equation}\label{SigmaE}
\Si(E)\!=\!\begin{dcases}
\scalebox{0.9}{
$\!-\tfrac{1}{\sqrt{E^2\!-\!4\ka^2}}\left[1\!-\! \left(\tfrac{E\!+\!\sqrt{E^2\!-\!4\ka^2}}{2\ka}\right)^{2l}\right]$},&\!E\!<\!-\!2\ka,\\
0,&\!E=\!-\!2\ka\cos\tfrac{\pi q}{l},\\
\scalebox{0.9}{
$\tfrac{1}{\sqrt{E^2\!-\!4\ka^2}}\left[1\!-\! \left(\tfrac{E\!-\!\sqrt{E^2\!-\!4\ka^2}}{2\ka}\right)^{2l}\right]$},&\!E\!>\!2\ka.
\end{dcases}
\end{equation}
Correspondingly, the energy shift  $\De(E)$ and the resonance width $\Ga(E)$ defined in Eqs.~(\ref{De}) and (\ref{Ga}) become
\begin{align}
\De(E)={}&\dfrac{\sin \left(2l\arccos\tfrac{E}{2\ka}\right)}{\sqrt{4\ka^2-E^2}}, \notag\\
\Ga(E)={}&\dfrac{1-\cos\left(2l\arccos\tfrac{E}{2\ka}\right)}{\sqrt{4\ka^2-E^2}}.
\end{align}
Following a similar calculation, Eqs.~(\ref{K}) and (\ref{I}) yield
\begin{equation}
K(z)=-\frac{\xi^2 \sin \left[N \arccos\left(-\frac{z}{2\lam}\right)\right]} {\lam\sin\left[(N+1) \arccos\left(-\frac{z}{2\lam}\right)\right]}
\end{equation}
and
\begin{equation}
I(z)=-\frac{\xi \sin \left[\arccos\left(-\frac{z}{2\lam}\right)\right]} {\lam\sin\left[(N+1) \arccos\left(-\frac{z}{2\lam}\right)\right]}
\end{equation}
in the complex plane. Projecting onto the real energy domain, the above expressions reduce to
\begin{equation}\label{KE}
K(E)=
\begin{dcases}
-\tfrac{\xi^2\sinh \left[N{\rm arccosh}\left(-\tfrac{E}{2\lam}\right)\right]} {\lam\sinh\left[(N+1){\rm arccosh}\left(-\tfrac{E}{2\lam}\right)\right]},& E\!<\!-2\lam,\\
\tfrac{\xi^2\sin \left(N\arccos\tfrac{E}{2\lam}\right)} {\lam\sin\left[(N+1)\arccos\tfrac{E}{2\lam}\right]},&| E|\!\le\! 2\lam,\\
\tfrac{\xi^2\sinh \left(N{\rm arccosh}\tfrac{E}{2\lam}\right)} {\lam\sinh\left[(N+1){\rm arccosh}\tfrac{E}{2\lam}\right]},& E\!>\!2\lam,
\end{dcases}
\end{equation}
and
\begin{equation}
I(E)=
\begin{dcases}
-\tfrac{\xi\sinh \left[{\rm arccosh}\left(-\tfrac{E}{2\lam}\right)\right]} {\lam\sinh\left[(N+1){\rm arccosh}\left(-\tfrac{E}{2\lam}\right)\right]},& E\!<\!-2\lam,\\
(-1)^{N+1}\tfrac{\xi \sin \left(\arccos\tfrac{E}{2\lam}\right)} {\lam \sin\left[(N+1)\arccos\tfrac{E}{2\lam}\right]},& | E| \!\le\! 2\lam,\\
(-1)^{N+1}\tfrac{\xi\sinh \left({\rm arccosh}\tfrac{E}{2\lam}\right)} {\lam\sinh\left[(N+1){\rm arccosh}\tfrac{E}{2\lam}\right]},& E\!>\!2\lam.
\end{dcases}
\end{equation}

\begin{figure}[htbp]
\centering
\includegraphics[width=\linewidth]{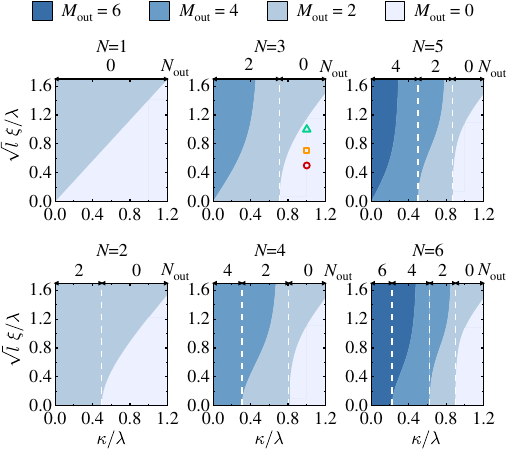}
\caption{Number $M_{\rm out}$ of BOCs of the tight-binding model for $N=1$ to $N=6$. $N_{\rm out}$ denotes the number of discrete energy levels lying outside the continuum, as determined by $N$ and $\ka/\lam$. The open red circle, open orange square, and open green triangle mark the parameter set $\kappa/\lambda = 1$, $\xi/\lambda = 0.5$ employed in Fig.~\ref{sp} for the case $N = 3$, corresponding to $l = 1$, $l = 2$ and $l = 4$, respectively.}
\label{M}
\end{figure}

We now specialize the general results for the count of BOCs, derived in  Sec.~\ref{sec:BOC}, to the present model. Since the discrete levels $\vep_n$ are symmetric about zero energy, the number $N_{\rm out}=2N_{\rm low}=2N_{\rm up}$ (with $0\le N_{\rm out}\le N$) of $\vep_n$ outside the continuum is
\begin{equation}
N_{\rm out}=
\begin{dcases}
2\left\lfloor \tfrac{N+1}{\pi} \arccos\tfrac{\ka}{\lam} \right\rfloor, & 0<\tfrac{\ka}{\lam}\le 1,\\
0, & \tfrac{\ka}{\lam}> 1,
\end{dcases}
\end{equation}
where $\lfloor\cdot\rfloor$ denotes the floor function. Given the self-energies at the band edges, $\Si(-2\ka)=- l/\ka$ and $\Si(2\ka)=l/\ka$, the two criteria for additional solutions below and above the band, namely $K(-2\ka)\le \Si^{-1}(-2\ka)$ and $K(2\ka)\ge \Si^{-1}(2\ka)$, merge into a single condition,
\begin{equation}\label{con2}
\frac{\sqrt{l}\xi}{\lam} \ge
\begin{dcases}
\sqrt{\tfrac{\ka\sin\left[(N+1)\arccos\tfrac{\ka}{\lam}\right]} {\lam\sin\left(N\arccos\tfrac{\ka}{\lam}\right)}},
& 0<\tfrac{\ka}{\lam}\le 1,\\
\sqrt{\tfrac{\ka\sinh\left[(N+1){\rm arccosh}\tfrac{\ka}{\lam}\right]} {\lam\sinh\left(N{\rm arccosh}\tfrac{\ka}{\lam}\right)}},
& \tfrac{\ka}{\lam}> 1,
\end{dcases}
\end{equation}
requiring the radicand to be nonnegative.
Figure~\ref{M} presents the number $M_{\rm out}$ of BOCs as a function of the chain length $N$. Owing to the spectral symmetries of the atomic chain and the photonic lattice, the total system hosts an even number of BOCs. The maximum possible count is $N+1$ for odd $N$ and $N$ for even $N$.

According to the results of Sec.~\ref{sec:BIC} and Eqs.~(\ref{Jw}) and (\ref{SigmaE}), the BICs that appear in this model all have zero self-energy and exist at energy $\ti{\vep}$ for any $l\geq 2$, provided that the discrete level at $\ti{\vep}$ satisfies the following condition:
\begin{equation}\label{BICcon}
\ti{\vep}=-2\ka\cos\frac{\pi q}{l}\quad (q=1,2,\ldots,l-1).
\end{equation}
Two special cases are particularly noteworthy. (i) For even $l$ and odd $N$, a BIC always exists at the zero-energy mode $\vep_{(N+1)/2}=0$, independent of the coupling strength. (ii) When $\ka/\lam=1$ and $l=N+1$, all $N$ discrete states associated with energy levels $\{\vep_n\}_{n=1}^N$ form BICs.

It is worth considering the limit $l\to\infty$, which corresponds to an infinite photonic lattice. In this case, the energy dispersion remains Eq.~(\ref{dispersion}) over $-\pi< k<\pi$, but the spectral density now changes to~\cite{Mahan2000,Economou2006,Garmon2009,Lombardo2014,Sanchez2017}
\begin{equation}
J(\om)=\frac{\Theta(4\ka^2-\om^2)}{\pi\sqrt{4\ka^2-\om^2}},
\end{equation}
showing van Hove singularities at the band edges $\om=\pm 2\ka$. We can also derive the self-energy
\begin{equation}
\Si(E)
=\begin{dcases}
-\tfrac{1}{\sqrt{E^2-4\ka^2}},& E<- 2\ka,\\
\tfrac{1}{\sqrt{E^2-4\ka^2}},& E>2\ka,
\end{dcases}
\end{equation}
vanishing energy shift $\De(E)$, and the resonance width
\begin{equation}
\Ga(E)=\frac{1}{\sqrt{4\ka^2-E^2}}.
\end{equation}
Since the condition~(\ref{con2}) always holds for $l\to \infty$ in each interval, the number of BOCs becomes independent of $\xi/\lam$ and is determined solely by $\ka/\lam$:
\begin{equation}
M_{\rm out}=
\begin{dcases}
N_{\rm out}+2, &\text{\parbox{6cm}{$\cos\tfrac{\pi(N_{\text{out}}+2)}{2(N+1)} < \tfrac{\kappa}{\lambda} \le \cos\tfrac{\pi N_{\text{out}}}{2N},$\\ and $0\le N_{\rm out}\le N-2$,}}\\[2ex]
N_{\rm out}, &\cos\tfrac{\pi N_{\text{out}}}{2N}<\tfrac{\kappa}{\lambda}\le \cos\tfrac{\pi N_{\text{out}}}{2(N+1)},\\
2,& \tfrac{\ka}{\lam}\ge 1.
\end{dcases}
\end{equation}
It is noted that there are at least two BOCs for any $N$ when the photonic lattice is infinite. Meanwhile, The violation of the condition $J(E)=0$ within the band precludes the existence of BICs.

\begin{figure}[htbp]
\centering
\includegraphics[width=\linewidth]{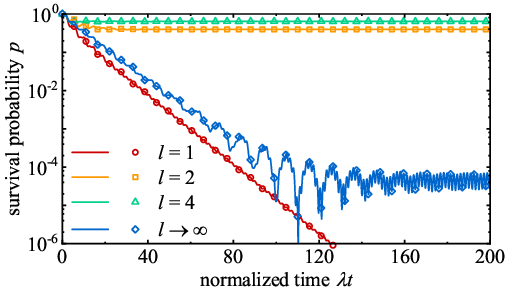}
\caption{Decay dynamics of the survival probability $p(t)$ for parameters $N=3$, $\ka/\lam=1$, and $\xi/\lam=0.5$. The results of Eq.~(\ref{pt}) (solid lines) and simulation (open symbols) are shown for four model geometries: coupling site $l=1$ (red line and circles), $l=2$ (orange line and squares), $l=4$ (green line and triangles), and the infinite photonic lattice limit $l\to\infty$ (blue line and diamonds).}
\label{sp}
\end{figure}

To check the time evolution of the system, we perform direct numerical simulations in the Wannier representation. As for single-excitation, we can expand the time-dependent state $|\phi(t)\ra$ as
\begin{equation}\label{abt}
|\phi(t)\ra=\sum_{\mu=1}^N \al_\mu(t)|\mu\ra_{\rm a}+\sum_{\nu=1}^\infty \be_\nu(t)|\nu\ra_{\rm p}.
\end{equation}
Substituting the ansatz Eq.~(\ref{abt}) and the model Hamiltonian~(\ref{TBMH}) into the time-dependent Schr{\"o}dinger equation, $\im |\dot{\phi}(t)\ra=\hH |\phi(t)\ra$, we obtain a set of coupled differential equations for the occupation amplitudes $\al_\mu(t)$ and $\be_\nu(t)$:
\begin{equation}\label{ns}
\begin{aligned}
\im\dot{\al}_N(t)={}&\!-\!\lam\al_{N\!-\!1}(t),\\
\im\dot{\al}_\mu(t)={}&\!-\!\lam[\al_{\mu\!-\!1}(t)\!+\!\al_{\mu\!+\!1}(t)], && 1\!< \!\mu \!<\! N,\\
\im\dot{\al}_1(t)={}&\!-\!\lam\al_2(t)\!+\!\xi\be_l(t),\\
\im\dot{\be}_\nu(t)={}&\!-\!\ka[\be_{\nu\!-\!1}(t)\!+\!\be_{\nu\!+\!1}(t)]\!+\!\de_{l\nu}\xi\al_1(t), && \nu \!>\! 1,\\
\im\dot{\be}_1(t)={}&\!-\!\ka\be_2(t)\!+\!\de_{l1}\xi\al_1(t).
\end{aligned}
\end{equation}
Starting from the initial state~(\ref{ini}), the survival probability can be calculated by $p(t)=\sum_{\mu=1}^N | \al_\mu(t) |^2$. In the simulation, we choose a large enough truncation ($10^3$ lattice sites) of the photonic lattice to avoid spurious reflections.

As an illustrative example, Fig.~\ref{sp} shows the decay dynamics of the survival probability $p(t)$ for a three-site chain $N=3$. Different coupling positions to the photonic lattice lead to three distinct decay regimes: a complete decay, a fractional decay, and an asymptotically oscillatory decay. We observe a perfect match between the solutions Eq.~(\ref{pt}) and the numerical simulations based on Eqs.~(\ref{ns}). For the coupling site $l=1$ and parameters $\ka/\lam=1$ and $\xi/\lam=0.5$, no bound states form, either outside or inside the continuum. This absence is confirmed by Fig.~\ref{M} for BOCs and by Eq.~\eqref{BICcon} for BICs. Therefore, the decay dynamics is determined by the time evolution of the scattering states, and the excitation undergoes an approximately exponential decay into the photonic lattice.
For the coupling site $l=2$, while no BOCs are formed either (see Fig.~\ref{M}), a BIC emerges at the energy $\vep_2=0$, in accordance with condition~(\ref{BICcon}), corresponding to special BIC case (i) above. This BIC localizes approximately half of the initial excitation within the atomic chain, leading to a survival probability that saturates at a finite steady value.
Similarly, as in the previous case, no BOCs are formed for the coupling site $l=4$ (see Fig.~\ref{M}), while all three discrete eigenstates become BICs at the energies $\vep=0,\pm\sqrt{2}\lam$, which is special BIC case (ii) above. Since each BIC is associated with a distinct initial eigenstate, they do not interfere with one another. As a result, the survival probability again saturates at a finite steady value, which is now larger than that for $l=2$ due to the increased number of contributing BICs.
A different picture can be found for the infinite photonic lattice limit $l\to\infty$. Here, two BOCs are present, but no BIC exists. Following an initial approximately exponential decay accompanied by oscillations, the survival probability asymptotically approaches low-amplitude oscillations around a small mean value.

\begin{figure}[htbp]
\centering
\includegraphics[width=\linewidth]{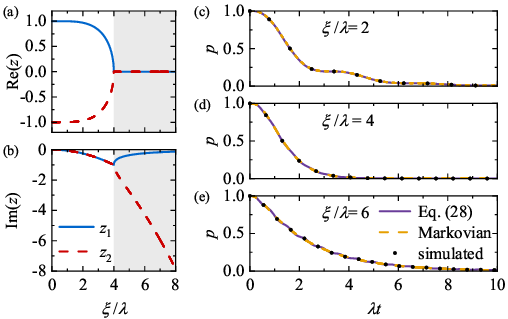}
\caption{Complex eigenvalues and survival probability dynamics. (a,b) Real and imaginary parts, $\Re(z_i)$ and $\Im(z_i)$, of the two eigenvalues of the effective Hamiltonian as functions of the coupling strength $\xi/\lam$ for $N=2$ and $\ka/\lam=4$. Blue solid and red dashed lines correspond to $i=1$ and $i=2$, respectively. The shaded region for $\xi/\lam >4$ indicates the $\mathcal{PT}$-symmetry-broken phase. (c)--(e) Survival probability $p(t)$ versus time $t$ for three representative couplings: $\xi/\lam=2$ ($\mathcal{PT}$-symmetric phase), $\xi/\lam=4$ (EP), and $\xi/\lam=6$ ($\mathcal{PT}$-symmetry-broken phase). The solid purple line, dashed orange line and solid black circle represent the Eq.~(\ref{pt}), Markovian approximate, and simulated results, respectively.}
\label{Markov}
\end{figure}

The present system also provides a promising testbed to explore non-Hermitian quantum phenomena~\cite{Ashida2020,Ding2022}. Under the Markovian approximation, e.g., $\ka/\lam\gg 1$, the energy-independent effective Hamiltonian~(\ref{HeffMar}) of a pair of coupled atoms with one attached to an infinite photonic lattice yields
\begin{equation}\label{H2}
\hH_{\rm eff}
=\begin{pmatrix}
-\lam-\im \dfrac{\xi^2}{4\ka} & -\im \dfrac{\xi^2}{4\ka}\\[2ex]
-\im \dfrac{\xi^2}{4\ka}   & \lam-\im\dfrac{\xi^2}{4\ka}
\end{pmatrix},
\end{equation}
which is anti-parity-time ($\mathcal{PT}$)-symmetric, i.e., $(\mathcal{PT})\hH_{\rm eff}(\mathcal{PT})^{-1}=-\hH_{\rm eff}$ \cite{Wu2014,Peng2016,Yang2017,Choi2018,Bian2023}.
The real and imaginary parts of the two complex eigenvalues of the Hamiltonian~(\ref{H2})
\begin{equation}
z_{1,2}=\pm\sqrt{\lam^2 - \left(\frac{\xi^2}{4\ka}\right)^2}-\im\frac{\xi^2}{4\ka},
\end{equation}
are plotted in Figs.~\ref{Markov} (a) and (b), respectively, for the corresponding eigenstates
\begin{equation}
| \Psi_{1,2}^+\ra=\left( \mp\sqrt{1-\frac{4\ka\lam}{\xi^2}}-\im\frac{4\ka\lam}{\xi^2},1\right)^{\rm T}.
\end{equation}
When $\xi^2<4\ka\lam$, the two complex eigenvalues satisfy $z_{1,2}^*=-z_{2,1}$. Moreover,  the corresponding right eigenstates $| \Psi_{1,2}^+\ra$ are also eigenstates of the $\mathcal{PT}$ operator, fulfilling $\mathcal{PT}| \Psi_{1,2}^+\ra=(\mp\sqrt{1-4\ka\lam/\xi^2}-\im 4\ka\lam/\xi^2)| \Psi_{1,2}^+\ra$, which confirms the system is in the $\mathcal{PT}$-symmetric phase.
In contrast, the $\mathcal{PT}$-symmetry of the states $| \Psi_{1,2}^+\ra$ is broken in the regime $\xi^2>4\ka\lam$. The $\mathcal{PT}$ operator now exchanges the two eigenstates,
$\mathcal{PT}| \Psi_{1,2}^+\ra=(\pm\sqrt{1-4\ka\lam/\xi^2}-\im 4\ka\lam/\xi^2)| \Psi_{2,1}^+\ra$, and the corresponding eigenvalues become purely imaginary.
The EP occurs at the critical coupling $\xi^2=4\ka\lam$, where the two eigenstates coalesce into a degenerate self-orthogonal state $| \Psi_{{\rm d},1}^+\ra=\left( -\im,1\right)^{\rm T}$.

The inevitable decay of the survival probability discussed in Sec.~\ref{sec:Mar} is exemplified in Figs.~\ref{Markov}(c) to \ref{Markov}(e). An excellent agreement among the solutions~(\ref{pt}), the Markovian approximate expressions~(\ref{decay}) and (\ref{ddcay}), and the simulated results based on Eq.~(\ref{ns}) confirms the validity of the Markovian approximation.
The decay behavior changes characteristically across different phases. In the $\mathcal{PT}$-symmetric phase, the two resonance states exhibit energy-level repulsion but share a common decay rate, resulting in an underdamped decay. In contrast, within the $\mathcal{PT}$-symmetry-broken phase, their energies attract while their decay widths bifurcate, leading to a double-exponential decay. Finally, at the EP, the coalescence of the two resonance states into a single degenerate state gives rise to an anomalous power-law exponential decay $p(t)=(2\lam^2 t^2+2\lam t+1)\e^{-2\lam t}$.

\section{Conclusion}\label{conc}

This paper presents an $N$-level Friedrichs model in which the interaction between multiple discrete states and a continuum is factorizable. The separable coupling condition arises naturally in various physical scenarios, including kaon phenomenology~\cite{Courbage2007}, a cesium atom coupled to a nanofiber~\cite{LeKien2005}, and atoms in photonic crystals~\cite{Cui2018,Shen2019,Burgess2022}.
Compared to the general model, it yields more concise eigenstate forms, but it does not improve the analytic solvability. However, this condition allows us to determine the number of BOCs via a simple graphical method, without numerically solving the full eigenvalue problem.
Criteria are established for counting BOCs by comparing $K(E)$ with the inverse of the self-energy $\Si(E)^{-1}$ at the continuum edge.
We analyze the decay of the survival probability and show that its long-time behavior falls into three regimes: complete decay with no bound states, saturation to a finite constant for a single bound state or for multiple BICs with zero self-energy, and persistent oscillations otherwise.
In the Markovian limit, the survival probability yields a non-exponential decay due to multi-resonance interference. At long times, excitations decay exponentially for a non-degenerate effective Hamiltonian, while a power-law exponential decay emerges at degeneracy.
Finally, we map a two-level atomic chain side-coupled to a photonic lattice onto the $N$-level Friedrichs model, and demonstrate the three long-time dynamic regimes outlined above through both theoretical derivations and numerical simulations. By constructing an anti-$\mathcal{PT}$-symmetric Hamiltonian on this platform, we explicitly demonstrate the distinct decay dynamics governed by different phases. A very recent experiment has emulated this model on a superconducting platform~\cite{Almanakly2026}.

What is particularly noteworthy is that the expressions for the bound and scattering states can also be obtained via the resolvent approach, which analyzes poles and branch cuts in the complex energy plane, or through the Fano diagonalization procedure~\cite{Fano1961,Longhi2009}.
In this work, we employ the Feshbach projection operator formalism to isolate the discrete subsystem of interest. Within this framework, an energy-dependent effective Hamiltonian  captures the discrete-subspace dynamics under continuum coupling. This approach offers two key advantages: (i) bound states and scattering states residing in the discrete subsystem correspond, respectively, to Hermitian and non-Hermitian effective Hamiltonians; (ii) it provides a natural pathway from a fully Hermitian description of the total system to a non-Hermitian open quantum system.
We believe this paper establishes a theoretical foundation for exploring dynamical properties and diverse applications of the Friedrichs model, thereby paving the way for future studies in a variety of contexts.
In particular, future work could be devoted to a systematic analysis of the detailed dynamics of related models via Green's function pole analysis, as demonstrated in Ref.~\cite{Dente2008}.

\section*{Acknowledgments}

We acknowledge grant support from the National Natural Science Foundation of China (Grants No. 12475024) and the Shandong Provincial Natural Science Foundation, China (Grants No. ZR2020QA079 and No. ZR2021MA081).

\bibliography{Friedrichs}

\end{document}